\documentclass[sigconf]{acmart}
\usepackage{url}            
\usepackage{amsfonts}       
\usepackage{nicefrac}       
\usepackage{microtype}      
\usepackage{graphicx}
\usepackage{amsmath}

\usepackage{amsthm}
\usepackage{subcaption}
\usepackage{amssymb}
\usepackage{subcaption}

\setcopyright{acmcopyright}
\copyrightyear{2018}
\acmYear{2018}
\acmDOI{10.1145/1122445.1122456}

\begin{document}

\title{Mitigating Bias in Online Microfinance Platforms: A Case Study on Kiva.org}

\author{Soumajyoti Sarkar}
\email{ssarka18@asu.edu}
\affiliation{%
  \institution{Arizona State University}
  \city{Tempe}
  \country{USA}
}

\author{Hamidreza Alvari}
\email{halvari@asu.edu}
\affiliation{%
  \institution{Arizona State University}
  \city{Tempe}
  \country{USA}
}







\renewcommand{\shortauthors}{Sarkar and Alvari}

\begin{abstract}
 Over the last couple of decades in the lending industry, financial disintermediation has occurred on a global scale. Traditionally, even for small supply of funds, banks would act as the conduit between the funds and the borrowers. It has now been possible to overcome some of the obstacles associated with such supply of funds with the advent of online platforms like Kiva, Prosper, LendingClub. Kiva for example, works with Micro Finance Institutions (MFIs) in developing countries to build Internet profiles of borrowers with a brief biography, loan requested, loan term, and purpose. Kiva, in particular, allows lenders to fund projects in different sectors through group or individual funding. Traditional research studies have investigated various factors behind lender preferences purely from the perspective of loan attributes and only until recently have some cross-country cultural preferences been investigated. In this paper, we investigate lender perceptions of economic factors of the borrower countries in relation to their preferences towards loans associated with different sectors. We find that the influence from economic factors and loan attributes can have substantially different roles to play for different sectors in achieving faster funding. We formally investigate and quantify the hidden biases prevalent in different loan sectors using recent tools from causal inference and regression models that rely on Bayesian variable selection methods. We then extend these models to incorporate fairness constraints based on our empirical analysis and find that such models can still achieve near comparable results with respect to baseline regression models. 
\end{abstract}


\maketitle

\section{Introduction}
Online lending in recent years has been considered to be an important contributor to financial restructuring in developing and underdeveloped nations by way of opening access to alternate sources of funding for them \cite{banerjee2015miracle}. Online platforms that enable such peer-to-peer transactions whereby certain groups of people invest in projects from poor entrepreneurs, have become very popular. These online platforms thus form an integral component of the credit disbursement process that was envisioned for microfinance. There exist different types of microlending services including for-profit lending services like LendingClub, Prosper and the pro-social platforms like Kiva\footnote{http://www.kiva.org}  where the lenders offer interest-free money to the borrowers. Platforms like Kiva are beneficial to borrowers, since lenders typically are risk-free indicating they do not expect any interest returns for the loan and hence can select their portfolio being less biased. Additionally, such pro-social platforms overcome the biases in loan disbursement through auctions in online platforms which is unfavorably inclined towards the credit-trustworthy users and undermines new users \cite{chen2014auctions}.

There have been a few studies conducted that debate whether such platforms have been successful in reaching countries where it would otherwise have been difficult to execute such peer lending strategies in an offline manner \cite{alfaro2008doesn, singh2018peer}. Successful mechanisms for peer lending come from both sides: the lender getting its expected return and the borrower finishing the project and repaying the loan. And to enable such successful mechanisms for lending, there has been considerable research in machine learning models that recommend projects to lenders \cite{choo2014understanding, rakesh2016probabilistic}, and to predict the dynamics of the crowdfunding \cite{zhao2017tracking}. This makes it imperative to understand whether there exists any bias when it comes to lenders selecting loans based on certain observable attributes and in presence of externalities like lender's perception of the economy.

Broadly, there have been a few groups of research studies conducted on understanding and promoting microfinance lending on such platforms. (1) Investigating biases: previous studies have focused on understanding and predicting bilateral trade transactions based on migration and GDP differences between country pairs. The goal here has been to investigate the presence of lender level preferences towards countries \cite{singh2018peer}. (2) Borrower and lender features: past studies include understanding various platform-external lender and borrower personal and regional characteristics that facilitate the transactions between countries \cite{paruthi2016peer, choo2014understanding} and the role of matching characteristics. However, the loan attribute concerning the loan sector is often overlooked especially to its connections to philanthropic and pro-social motivations of investors. Such detailed connections of the loan sector on funding propensity have been investigated in \cite{ly2010individual}, (3) Fairness aware lending: recent studies have acknowledged the existence of bias in lending models and the need to diversify the distribution of donations to reduce the inequality of loans \cite{lee2014fairness}, and  (4) Social networks: the role of networks have been studied from the perspective of facilitating bidding behavior in platforms where the investors stand to profit from their lending \cite{lin2013judging}. Similarly, teams and group loans have been understood to have a more positive effect on gaining traction from lenders as have been studied in \cite{pham2017deep}.

What is often overlooked is the influence of external factors pertaining to the borrower countries that influence lender preferences and which cannot be directly observed from the platform data. Furthermore, there has been substantial evidence in the recent past that supports Lucas paradox, which indicates that, counter-intuitively the liberalization of international capital regimes using the internet platforms has not produced an open club, rather a rich club, a group of countries that exhibit the country-pair bias \cite{alfaro2008doesn}. Since recommendation models typically do not consider such external data while building their models \cite{rakesh2016probabilistic}, such latent biases arising from external factors including lender perceptions of countries\footnote{https://bit.ly/2LF9Mpp} can be quite detrimental for certain projects especially ones from specific countries. 

To this end, we investigate the factors behind the funding speed of loans using the dataset available from Kiva. The goal is to see whether the lenders fall for region specific economic factors that they expect would help them avoid loan defaults from borrowers and whether that affects funding projects in certain sectors. We compare the effect of different sectors on project funding times when the economic external factors form part of the models in consideration. Using data from 143,856 loans over a period of 4 years and economic indicators from World Bank Data, we make the following contributions:

\begin{itemize}
   \item We gather data from Kiva loans and heterogeneous data sources and build regression models to estimate the impact of such factors on the funding speed. We observe the role of the project or loan sector as a sensitive attribute in the models especially when its correlation with the funding speed differs for different sectors.
    
    \item We use recent causal inference and machine learning tools to estimate the effects the sector attribute on funding times. We specifically find that loans catering to Retail are funded 4 days slower relative to the other sectors on aggregate and loans for Arts are funded 6 days faster - all these when considering the economic factors of the location of the borrowers and the loan attributes. This is in contrast to observations from data that do not reveal such hidden discrepancies when excluding external factors.
    
    \item Following this, we incorporate fairness driven constraints to mitigate some of the  biases arising from these loan specific attributes for particular sectors of loans. Our results suggest that even with such fairness constraints, the model performances are not too far-off from baselines, thus giving hope for future systems that take into account such constraints.
\end{itemize}

We note that this is the first work in attempting to understand the existing biases from loan attributes when external factors are also considered to be the contributors to such decisive disparities. Throughout our work, we mainly focus on linear regression models, however we adopt the models to use Bayesian variable selection techniques that not only fits the regression models while simultaneously picking the most effective regressors, but also allows us to incorporate any prior beliefs about the attributes of the projects and the users that cannot generally be accomplished with standard regression models. 

\begin{figure}[!t]
\centering
\minipage{0.25\textwidth}
\includegraphics[width=4.5cm, height=3cm]{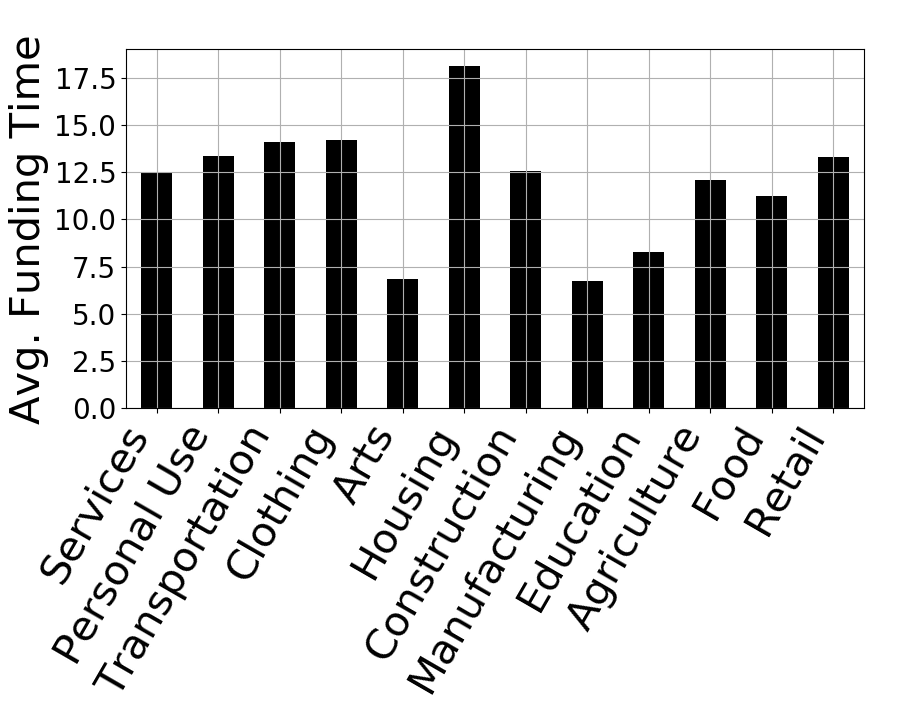}
\subcaption{}
\endminipage
\minipage{0.3\textwidth}
\includegraphics[width=4.5cm, height=3cm]{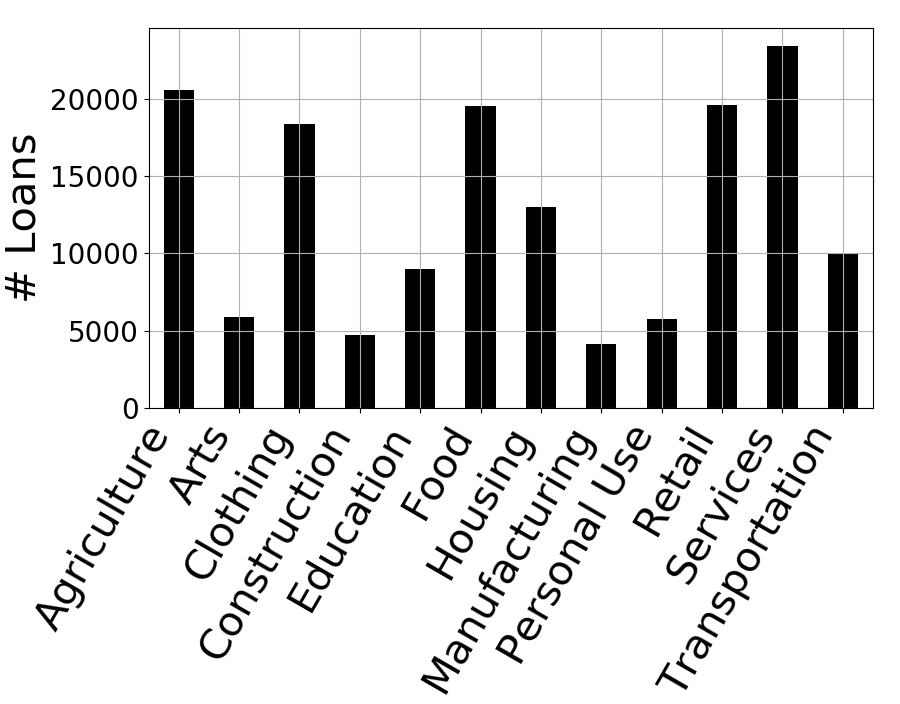}
\subcaption{}
\endminipage
\hfill
\\
\minipage{0.25\textwidth}
\includegraphics[width=4.5cm, height=3cm]{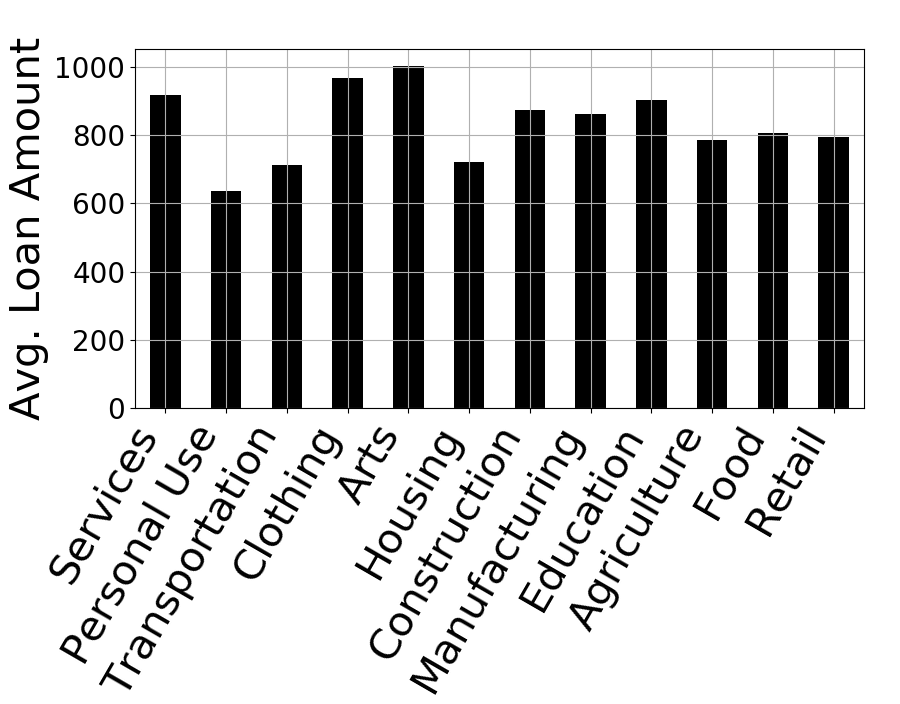}
\subcaption{}
\endminipage
\hfill
\minipage{0.2\textwidth}
\includegraphics[width=4.1cm, height=3cm]{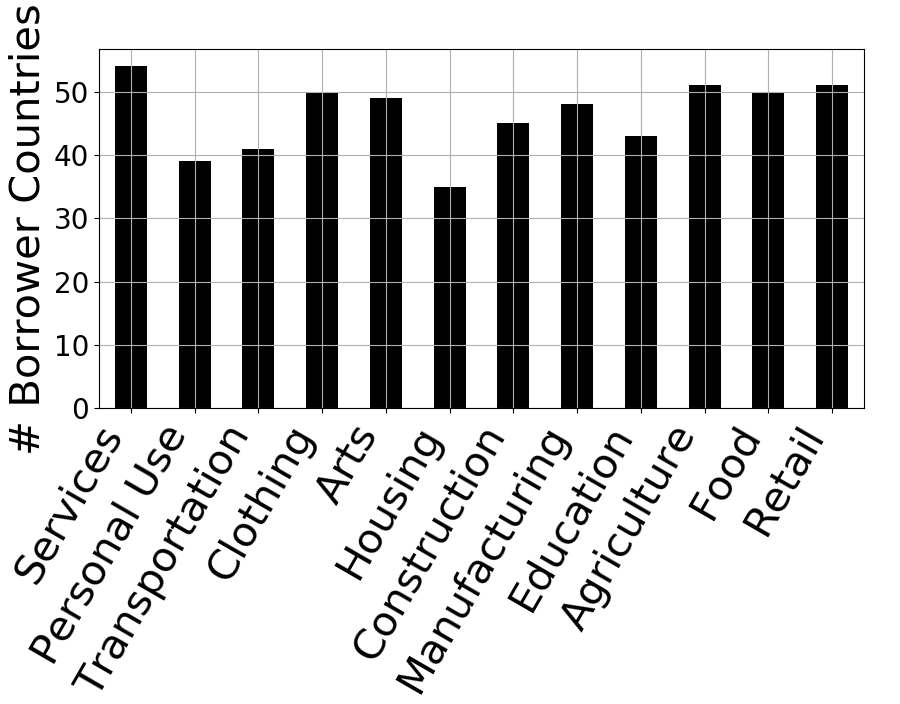}
\subcaption{}
\endminipage
\hfill
\\
\caption{Distribution of (a) average funding times, (b) number of loans, (c) average loan amount by loan sectors and (d) number of borrower countries by each sector.}
\label{fig:loans_stats}
\end{figure}
\section{Data}
Kiva is a non-profit micro-financial organization and its lending model is based on crowdfunding in which any individual can fund a particular loan by contributing to a loan individually or as a part of a lender team. The choice behind this platform is driven by the motivation to test a few hypotheses in this research - we want to be able to understand the presence or absence of behavioral and social bias that could create preferences for certain projects. Since public perceptions of societies can elicit biases towards countries with specific geographical, cultural or political fabric and that can affect funding in such online platforms, we set out to test the interplay of economic externalities and the loan specific attributes in such settings.

The publicly available Kiva dataset\footnote{https://www.kiva.org/build/data-snapshots} contains various entities: (1) the data for the loans that contains various attributes associated with the borrowers, (2) the lenders' information containing various attributes regarding a lender's history of funding projects (3) the borrowers' information containing various attributes regarding a borrower's project and repayment history, (4) field partner which acts as the mediator and allocates loans from the lenders to the borrowers. Since our objective in this study is to understand the role of developmental factors when paired with the sector that receives the most funding, we use the following attributes that are associated with a loan in Kiva's platform from January 2010 to December 2014: (1) \underline{sector}: categorical attribute denoting the sector of loan activity. The sectors considered in our study after removing for sparse data is shown in Figure~\ref{fig:loans_stats}. Note that the set of sector tags are fixed for all loans and are not randomly generated. (2) \underline{currency policy} - binary attribute to reduce risk of currency fluctuation\footnote{https://pages.kiva.org/blog/new-kiva-feature-currency-risk-protection}, (3) \underline{language} - the language of the loan description - since 70\% of the loans we considered were in English, we converted this to a binary attribute by considering all non-English languages as one category, (4) \underline{loan amount} - numerical attribute denoting the amount of loan requested for the project, (5) \underline{borrower gender} - binary attribute denoting the gender of the borrower, and (6) \underline{funding time} - this is a derived numerical attribute  calculated as the difference between the time of the loan request and the time when it was fully funded. We use this attribute for measuring the preference of the investors towards particular projects and our models are based on understanding what attributes account for lesser funding times. We plot the distribution of the funding times and the number of loans by sectors in our dataset in Figure~\ref{fig:loans_stats}. However, unlike similar analyses, we do not found any substantial difference in the average loan amounts by sectors that get funded. Apart from entertainment, most of the sectors have similar funding requests that ultimately get funded.

\begin{table}[!t]
\small
\begin{tabular}{|l|l|l|}
\hline
\textbf{\# Loans}     & \textbf{\# Lender Countries}    & \textbf{\# Borrower Countries} \\ \hline
143856                & 216                             & 57                             \\ \hline \hline
\textbf{\# Languages} & \textbf{Avg. Loan Amount (USD)} & \textbf{Avg. Funding Time (std)}     \\ \hline
7                     & 836.18                          & 12.58 (14.6)              \\ \hline
\end{tabular}
\caption{Basic statistics for loans used in our study} \label{tab:stats}
\vspace{-8mm}
\end{table}

As a first task, we try to investigate the causal effects of borrower-lender differences arising from lender perceptions of the borrower countries as well as implicit economic and cultural variations. We try to measure the extent of impact it has on the funding times when considered alongside the sector of the loans. To this end, for each loan, we gather the following data from the world bank metrics dating back to 2010 \cite{world2013world}. We gather the following attributes: (1) \underline{ease of business}: an ordinal attribute denoting the rank of the borrower country for ease of business, (2) \underline{loan access}  - numerical attribute denoting the ease of access to loan in the borrower country through formal financial institutions, (3) \underline{women ratio} - numerical attribute measuring the ratio of women in labor force compared to  men, (4) \underline{affordability} - numerical attribute pertaining to the costs associated with using services, including both interest rates and fees, (5) \underline{VC financing} - ordinal attribute that indicates how easy it is for the borrower to seek capital locally or otherwise in their country, (6) \underline{capacity innovation} - ordinal attribute denoting the capacity of people in the borrower country to innovate and (7) \underline{internet penetration} - numerical attribute denoting the percentage of people in the borrower country using the internet. To measure the cross-cultural similarities, we proceed as done in  \cite{singh2018peer} to use the following features: (8) \underline{colonization} -  binary attribute denoting whether the borrower country was colonized by lender country, (9) \underline{distance} - geographical distance between borrower and lender countries obtained from \cite{mayer2011notes},  (10) \underline{migrants} - numerical attribute that measures the number of people or borrower country origin living in the lender country obtained from world bank data and  (11) \underline{GDP difference} - numerical attribute denoting the GDP Difference between borrower and lender countries obtained from world bank data.   For the derived attributes which were calculated based on borrower and lender countries, we used the following method: for numerical attributes, for a specific loan we took the average of all the borrower-lender pairs for that loan. For categorical or binary  attributes like colonization, we randomly picked one of the borrower-lender pairs for that loan and used that for the loan feature. This however introduces some approximation into the feature measurements. For all numerical attributes, we performed standardization for the regression models which would be described henceforth.

The dataset is publicly available for download\footnote{https://bit.ly/2TnqhL7}. After merging the data from these heterogeneous sources, we list the basic statistics of the loans used in this study and is shown in Table~\ref{tab:stats}. We find that while the average funding time is 12.5 days, the standard deviation is 14.6 days, which demands further investigation behind the variations. 


\begin{table}[!t] 
\small
\begin{tabular}{|l|l|l|l|l|l|} 
\hline
\textbf{OLS Estimates}                                        & \textbf{M1}               & \begin{tabular}[c]{@{}l@{}} \textbf{M2} \\ \textbf{ (Sector)} \end{tabular} & \begin{tabular}[c]{@{}l@{}} \textbf{M3} \\ \textbf{(Services)} \end{tabular}   & \begin{tabular}[c]{@{}l@{}} \textbf{M4} \\ \textbf{(Agri.)} \end{tabular} & \begin{tabular}[c]{@{}l@{}} \textbf{M5} \\ \textbf{(Retail)} \end{tabular}\\ \hline
{\color[HTML]{000000} \textbf{Intercept}}                     & {\color[HTML]{000000} 21.0346} & {\color[HTML]{000000} 20.2312}         & {\color[HTML]{000000} 22.3927} & {\color[HTML]{000000} 20.4209} & 19.9626 \\ \hline
\textbf{Sector}                                               & {\color[HTML]{000000} }        &                                        & {\color[HTML]{FE0000} -1.5268} & {\color[HTML]{FE0000} -0.5921} & {\color[HTML]{FE0000} 2.3072} \\ \hline
\begin{tabular}[c]{@{}l@{}} \textbf{Currency} \\ \textbf{Policy[T.shared]} \end{tabular} & {\color[HTML]{000000} -1.0349} & {\color[HTML]{000000} -1.1697}         & {\color[HTML]{000000} -3.2858} & {\color[HTML]{000000} -0.894} &  -0.2923 \\ \hline
{\color[HTML]{000000} \textbf{Language}}                      & {\color[HTML]{000000} -2.5225} & {\color[HTML]{000000} -2.0611}         & {\color[HTML]{000000} -1.3739} & {\color[HTML]{000000} -2.1189}  & -2.3513 \\ \hline
{\color[HTML]{000000} \textbf{Ease of business}}              & {\color[HTML]{000000} -0.0313} & {\color[HTML]{000000} -0.027}          & {\color[HTML]{000000} -0.0309} & {\color[HTML]{000000} -0.0282} & -0.0281  \\ \hline
{\color[HTML]{000000} \textbf{Colonization}}                  & {\color[HTML]{FE0000} -1.6048} & {\color[HTML]{FE0000} -2.0484}         & {\color[HTML]{FE0000} 1.5085}  & {\color[HTML]{FE0000} -3.7422} & {\color[HTML]{FE0000} -0.8401} \\ \hline
\begin{tabular}[c]{@{}l@{}} \textbf{Borrower Gender} \\ \textbf{[T.female]}  \end{tabular}          & {\color[HTML]{000000} -4.2708} & {\color[HTML]{000000} -4.609}          & {\color[HTML]{000000} -4.4875} & {\color[HTML]{000000} -4.1959} & -4.7179\\ \hline
{\color[HTML]{000000} \textbf{Loan Amount}}                   & {\color[HTML]{000000} 3.8751}  & {\color[HTML]{000000} 4.1365}          & {\color[HTML]{000000} 3.6231}  & {\color[HTML]{000000} 5.1564} &  3.6773\\ \hline
{\color[HTML]{000000} \textbf{Distance}}                      & {\color[HTML]{000000} -0.7032} & {\color[HTML]{000000} -0.6359}         & {\color[HTML]{000000} -0.6739} & {\color[HTML]{000000} -0.8225} & -0.3331 \\ \hline
{\color[HTML]{000000} \textbf{Migrants}}                      & {\color[HTML]{000000} -2.1397} & {\color[HTML]{000000} -2.4934}         & {\color[HTML]{000000} -1.9945} & {\color[HTML]{000000} -2.3797} & -2.4833\\ \hline
{\color[HTML]{000000} \textbf{GDP Difference}}                & {\color[HTML]{000000} -0.3162} & {\color[HTML]{000000} -0.2086}         & {\color[HTML]{000000} -0.0415} & {\color[HTML]{000000} -0.2439} & 0.0622\\ \hline
{\color[HTML]{000000} \textbf{Loan Access}}                   & {\color[HTML]{FE0000} -1.2037} & {\color[HTML]{FE0000} -0.526}          & {\color[HTML]{FE0000} 1.619}   & {\color[HTML]{FE0000} -1.582} & {\color[HTML]{FE0000} 0.2476} \\ \hline
{\color[HTML]{000000} \textbf{Women Ratio}}                   & {\color[HTML]{000000} -2.9224} & {\color[HTML]{000000} -3.2598}         & {\color[HTML]{000000} -2.1065} & {\color[HTML]{000000} -2.0469}  & -3.3524\\ \hline
{\color[HTML]{000000} \textbf{Affordability}}                 & {\color[HTML]{FE0000} -2.2799} & {\color[HTML]{FE0000} -2.2284}         & {\color[HTML]{FE0000} 2.4288}  & {\color[HTML]{FE0000} -2.2366} & {\color[HTML]{FE0000} -2.7703} \\ \hline
{\color[HTML]{000000} \textbf{VC finance}}                    & {\color[HTML]{000000} 2.2952}  & {\color[HTML]{000000} 2.0675}          & {\color[HTML]{000000} 0.4367}  & {\color[HTML]{000000} 2.3493}  & 1.042\\ \hline
{\color[HTML]{000000} \textbf{Capacity innov.}}               & {\color[HTML]{000000} -0.0378} & {\color[HTML]{000000} -0.0177}         & {\color[HTML]{000000} 0.5504}  & {\color[HTML]{000000} 0.5695} & -0.0052\\ \hline
{\color[HTML]{000000} \textbf{Internet Pen.}}          & {\color[HTML]{000000} -0.2639} & {\color[HTML]{000000} 0.4767}          & {\color[HTML]{000000} 0.7634}  & {\color[HTML]{000000} -1.3048} & 0.9828 \\ \hline
\end{tabular}
\hfill
\caption{Table: OLS Regression estimates on funding time for a project loan. For model \textbf{M1}, we do not include the Sector attribute and for model \textbf{M2}, the attribute Sector (categorical) is used as the dummy variable.}
\label{tab:regress_1}
\vspace{-8mm}
\end{table}

\section{Preliminary analysis of potential disparities}
We begin with  a simple linear regression model to investigate the importance that these economic indicators capturing the borrower's nations, have on the funding time and how they play a role compared to the loan sector. When we regress the variables of the economic factors and the loan attributes barring the loan sector, on the funding time denoted by model \textbf{M1}, we find from Table~\ref{tab:regress_1} instantly that there are some borrower-lender attributes that have a larger role to play - the distance, and GDP difference have a negligible impact on the funding time whereas the feature measuring the migrants of borrower country in lender country has a negative correlation with funding times - it suggests that the cultural similarities arising from cross-border migration results towards faster funding for borrowers with such cultural advantages. Similarly,  the women ratio factor has a significant negative correlation on the funding time along with the borrower gender in line with previous research \cite{alfaro2008doesn}.  This indicates that the perception about the role of women in such economies is a significant driver towards deciding whether the project would receive faster funding. This is also supported by the observation that the borrowers' gender shares such a correlation when regressed with the female category as the positive category in the binary attribute.

Next, we include the project sector attribute in the regression model as a dummy variable with the corresponding one-hot encodings denoted by model \textbf{M2}. We observe that while there are minor changes in the magnitude of the coefficients, the correlations of these variables do not change in the presence of the sector variable. This tempts us to conclude that when recommendation systems rely on these attributes to predict the best projects in terms of having better chances of funding, they can make a pretty fair classification for all projects based on these attributes. However, upon closer analyses, we look into the effect that these attributes have when considering each sector at a time. Taking each loan sector category $s$ as the main category in contention, we build sector specific models. For a model on sector $s$, we consider all loans belonging to $s$ as one category and all other categories as a unified dummy sector category separate from the sector $s$. We build regression models for all 12 sectors in a similar fashion. We show four of the models corresponding to four different sectors in Table~\ref{tab:regress_1} - in model \textbf{M3} we convert the multi-category sector attribute into a binary category by considering the services sector and loans belonging to that category as one cluster and all the other loans belonging to other sectors as the other cluster. When we compare M3 with model \textbf{M4} catering to the agriculture sector, we can observe that not only do some of the attributes differ in the magnitude significantly, but the influence from the attributes also reverses in some cases. Particularly, we find that the influence from the attribute colonization has opposite effects for the two sectors and similarly for attributes like affordability and loan access. 

When comparing the role of sectors, we observe that the coefficient magnitude demonstrates that the relative number of days by which each sector gets funded faster or slower relative to the other sectors. However,  the fact that these results are also heavily affected by the varying sparsity of the data. This leads us to turn our attention to recent literature on more robust causal reasoning tools that allow for explaining the effect of sectors on funding speed in the presence of such externalities \cite{pham2017deep, athey2017estimating}.  We will compare these results later with those obtained from causal reasoning based mechanisms and show how the disparity calls for controlling existing biases among the sectors.

\section{Causal inference} \label{sec:causal}
Note that the treatment of interest here is the loan sector assignment for the loan requested and we are interested in estimating the effects of loan sector relative to the economic, cultural and other loan characteristics, on the funding time. Following the work done in \cite{pham2017deep}, we would use the Robin Causal Model or the Potential Outcome Framework to estimate the treatment effects.

\subsection{Treatment Effects Indicators}

We describe in this section how we measure the causal impact metrics for each sector $s$. We estimate the treatment effects of sector on loan funding time considering separate models for each sector and treating the whole batch of data separately for each sector. Let the features be denoted by $X$, which in our case are all the attributes except the project sector $s$. Let $Y$ be the outcome of interest, in our case the funding time of the loans. For each sector $s$, we consider $W$ to be the binary treatment variable (whether a loan belongs to $s$ or not). Following this, for each sector $s$, we represent the dataset in the form $(Y_i, X_i, W_i)_{i=1}^n$, where $W_i$ denotes whether the sector for loan $i$ is $s$ or not ($W_1=1$ when loan belongs to $s$), $n$ denoting the number of loans in the data. Note that $W_i$ would be different for loan $i$ when considering different sectors since the observational data gives us the actual loan sector.  We will drop the subscripts from $W$ when generalizing the inference settings for all loans. We will also refrain from attaching $s$ as sub/super-scripts to notations since we perform all the following steps and estimate models in the same was irrespective of the sectors. We are interested in estimating the average treatment effects (ATE) of $W$ on $Y$ for each sector $s$ and this is given by:

\begin{equation}
    \tau = \mathbb{E}[Y(1) - Y(0)]
\end{equation}

\noindent where $Y(1)$ is the potential outcome of a loan that belongs to s while $Y(0)$ is the one that does not belong to s. However, in the data, only one of them is observed for each loan when considering models for a specific sector. The hree assumptions that are made during this estimation procedure are: (1) $(SUTVA)$ - The apriori assumption that the value of $Y_i$ when instance $i$  is exposed to treatment $W_i$  will be the same, no matter what mechanism is used to assign the treatment to $i$ and no matter what treatments others receive, (2) the probability of outcome $Y_i$ is independent of the features $X_i$ given $W_i$ - it means that the features $X_i$ do not simultaneously affect $W_i$ and $Y_i$. In our case this is more intuitive since firstly the external economic factors in itself have no bearing on the choice of the loan sectors and secondly, the loan sector also has little in relation to other loan features like gender, loan amount,  and (3) both treatment and control groups have has at least one instance assigned to them (see \cite{morgan2015counterfactuals, pham2017deep} for more details on these assumptions).

\subsection{Estimating Treatment Effects}
\label{sec:steps_causal}
With recent advances in machine learning to create estimators for ATE \cite{athey2016efficient, belloni2014inference, hill2011Bayesian}, we use the Doubly Robust Estimator (DRE) \cite{robins2000robust,kang2007demystifying} to measure $\tau$.  We briefly lay out the steps for estimating $\tau$ using DRE for our data - note we follow these steps for all sectors individually: 

\begin{enumerate}
    \item  \textbf{Outcome Model} - For loan sector $s$, we consider the loans $i$ belonging to $s$ as having $W_i$=1 and all other loans as $W_i$=0. Then we use the treated data $\{i:W_i=1\}$ to estimate $\mu(1, x)=\mathbb{E}[Y(1)|X=x]$ with estimator $\hat{\mu}(1, x)$ and use control data $\{i:W_i=0\}$ to estimate $\mu(0, x)=\mathbb{E}[Y(0)|X=x]$ with estimator $\hat{\mu}(0, x)$.
    
    \item \textbf{Propensity Score Model}: We then estimate the propensity score model - use all loans data to estimate $e(x)$ = $\mathbb{P}(W=1|X=x)$ with estimator $\hat{e}(x)$.
    
    \item The DRE $\hat{\tau}_{DRE}$ is given by 
        \begin{gather*}
        \hat{\tau}_{DRE} = \frac{1}{n}\sum_{i=1}^n\Big[\ W_i \times \frac{Y_i-\hat{\mu}(1, X_i)}{\hat{e}(X_i)} \\ - (1-W_i) \times \frac{Y_i - \hat{\mu}(0, X_i)}{1-\hat{e}(X_i)}    - \hat{\mu}(1, X_i) -  \hat{\mu}(0, X_i) \Big]
        \end{gather*}
    
    \item The standard error is then estimated following \cite{lunceford2004stratification} by using an empirical sandwich estimator. For each instance/loan $i$, we have 
    \begin{gather*}
        IC_i = W_i \times \frac{Y_i - \hat{\mu}(1, X_i)}{\hat{e}(X_i)} - (1-W_i) \times \frac{Y_i - \hat{\mu}(0, X_i)}{1 - \hat{e}(X_i)}  \\ + \hat{\mu}(1, X_i) - \hat{\mu}(0, X_i) - \hat{\tau}_{DRE}
    \end{gather*}
    and  $\sigma^2$ = $\frac{1}{n}\sum_{i=1}^n IC_i^2$. The standard  error is estimated as $\frac{\sigma}{\sqrt{n}}$. The 95\% confidence interval of $\tau$ is estimated by $(\hat{\tau}_{DRE} - z_{0.975}\frac{\sigma}{\sqrt{n}}, \hat{\tau}_{DRE} + z_{0.975}\frac{\sigma}{\sqrt{n}})$.
\end{enumerate}

The DRE has the double robustness property: given that either the outcome model or the propensity score model or both are correctly specified, the estimator is consistent. In the following subsections, we briefly outline the techniques we adopt for these 4 steps mentioned. 

\subsection{Learning Outcome Models} \label{sec:learning_model}
In order to estimate $\hat{\mu}(1, x)$ and $\hat{\mu}(0, x)$ for each sector $s$, we use regression models, however we observe from Table~\ref{tab:regress_1} that not all variables are equally important when measuring their outcome on funding times and these differ substantially among the sectors. To this end, we adopt some variable selection techniques while building separate regression models for  $\hat{\mu}(1, x)$ and $\hat{\mu}(0, x)$ for a sector $s$. We specifically adopt Bayesian methods  where  sparsity can be favored by assuming sparsity-enforcing priors on the model coefficients. These types of priors are characterized by density functions that are peaked at zero and also have a large probability mass in a wide range of non-zero values. Ideally, the posterior mean of truly zero coefficients should be shrunk towards zero. At the same time the posterior mean of non-zero coefficients should remain unaffected by the assumed prior. We use spike-and-slab priors which have some advantages when compared to other sparsity enforcing priors like Laplace and Student’s $t$ priors \cite{o2009review}. The advantage of Bayesian methods in such settings is that it allows us to also get credible intervals on the posterior estimates of the model coefficients and it allows us to easily incorporate some additional constraints in the priors which we will discuss in the later sections. We briefly review the spike-and-slab model \cite{ishwaran2005spike} as the regression model in choice and we learn separate models for $\hat{\mu}(1, x)$ and $\hat{\mu}(0, x)$ for a specific sector. 

Let $\mathbf{y} \in \mathbb{R}^{n \times 1}$ be an $n$-dimensional row vector denoting the target variable and $\mathbf{X} \in \mathbb{R}^{n \times p}$ denote the design matrix, $p$ denoting the number of attributes in our model except the sector attribute. Briefly, the spike-and-slab model specifies the prior hierarchy in the following way:

\begin{align}
\begin{split}
y_i \sim N(\beta x_i, \sigma^2) \\
\beta_i \sim (1-\pi_i)\delta_0 + \pi_i N(0, \sigma^2\tau^2) \\
\tau^2 \sim \mbox{Inverse-Gamma}(\frac{1}{2}, \frac{s^2}{2}) \\
\pi_i \sim \mbox{Bern}(\theta) \\
\theta \sim \mbox{Beta}(a,b) \\
\sigma^2 \sim \mbox{Inverse-Gamma}(\alpha_1, \alpha_2)
\end{split}
\label{eq:spike_slab}
\end{align}

\noindent where $i \in [1, p]$ indexes the features in the regression model, $\beta$ denotes the coefficients in the regression model. The first equation defines a regression model where the response $y_i$ follows a normal distribution conditioned on $\mathbf{x}_i$ and the parameters $\beta$. The normal distribution has a variance $\sigma^2$. The second equation models the way in which sparsity is enforced on the model coefficients. The sparsity of $\mathbf{\beta}$ can be favored by assuming a spike-and-slab prior for the components of this vector - the slab $N(0, \sigma^2\tau^2)$ is a zero mean broad Gaussian whose variance $\tau^2$ is large and the scale $\sigma^2$ is multiplied so that the prior scales with outcome. The spike $\delta_0$ is a Dirac Delta function (point probability mass) centered at 0 and this component is responsible for deciding whether the posterior for these coefficients would be zeroed out. $\pi \in [0,1]$ is a mixture weight between the spike-and-slab components in the prior. The rest of the equations denote the hierarchical structure of the parameters $\sigma^2$, $\tau^2$ and $\pi$. Note that $\tau^2$ and $\theta$ are common to all predictors .We now briefly describe how we sample the parameters using the Gibbs sampling technique for generating markov chain traces. The details of the sampling procedure have been added to the Appendix section.

\subsection*{Sampling $\theta$}
 To sample the conditional posterior $p(\theta|\mathbf{y}, \mathbf{\beta}, \mathbf{\pi}, \tau^2, \sigma^2)$= $p(\theta|\mathbf{\pi})$, we calculate the following as:
$p(\theta|\mathbf{\pi}) =  \frac{\theta^{a+\sum_{i=1}^p \pi_i-1}(1-\theta)^{b+\sum_{i=1}^n(1-\pi_i)-1}}{\int \theta^{a+\sum_{i=1}^p \pi_i-1}(1-\theta)^{b+\sum_{i=1}^n(1-\pi_i)-1} d\theta} $

\noindent where $B$ is the beta function. The posterior is $\theta|\mathbf{\pi}$ $\sim Beta\Big( a + \sum_{i=1}^p \pi_i, b + \sum_{i=1}^n (1-\pi_i) \Big)$.

\subsection*{Sampling $\tau^2$}
The conditional posterior of $\tau^2$ can be derived from the probability $p(\tau^2|\mathbf{y}, \mathbf{\beta}, \mathbf{\pi}, \theta, \sigma^2)$= $p(\tau^2|\mathbf{\pi}, \mathbf{\beta})$.

Here since $\pi$ can assume values 0 or 1 we tackle each case independently and derive the following. We sample from the prior if all $\pi_i$'s are zero. Let $\mathbf{\pi}$=$\{\pi_1, \ldots, \pi_p\}$ be the vector of mixture weights and let $\mathbf{0}$ be a vector of zeros of length $p$. Following this, we have

\begin{align} \label{eq:tau_post_2}
\begin{split}
p(\tau^2|\mathbf{\pi}, \mathbf{\beta}) &= \frac{1}{Z}p(\mathbf{\beta}|\tau^2, \mathbf{\pi})p(\pi)p(\tau^2) \\
                       &= \frac{1}{Z} \prod_{i=1}^p\pi_i (2\pi\sigma^2\tau^2)^{-\frac{1}{2}} exp\Big( -\frac{1}{2\sigma^2\tau^2} \mathbf{\beta}^T\mathbf{\beta} \Big) \\ & \frac{\Big( \frac{s^2}{2}\Big)^{\frac{1}{2}}}{\Gamma\Big( \frac{1}{2}\Big)} (\tau^2)^{-\frac{1}{2}-1} exp\Big(- \frac{\frac{s^2}{2}}{\tau^2}\Big) \\
\end{split}
\end{align}
\noindent which is a Gamma distribution and therefore we sample $\tau^2|\mathbf{\beta}, \mathbf{\pi}$ $\sim \mbox{Inverse-Gamma}(\frac{1}{2} + \frac{\sum_{i=1}^p\pi_i}{2}, \frac{s^2}{2} + \frac{\mathbf{\beta}^T\mathbf{\beta}}{2\sigma^2})$. On the other hand, when $\pi=0$, the $\beta_i$'s are 0 and we simply sample from the prior $\tau^2|\mathbf{\beta}, \mathbf{\pi}$ $\sim \mbox{Inverse-Gamma}(\frac{1}{2}, \frac{s^2}{2})$.

\subsection*{Sampling $\sigma^2$}
The conditional posterior of $\tau^2$ can be derived in a similar manner as above from the probability $p(\sigma^2|\mathbf{y}, \mathbf{\beta}, \mathbf{\pi}, \theta, \sigma^2)$= $p(\sigma^2|\mathbf{y}, \mathbf{\beta})$. Proceeding as before, we can derive the sampling as follows:  $\sigma^2|\mathbf{y}, \mathbf{\beta}$ $\sim \mbox{Gamma} \Big( \alpha_1 + \frac{n}{2} , \alpha_2 + \frac{(\mathbf{y} - \mathbf{X}\beta)^T(\mathbf{y} - \mathbf{X}\beta)}{2}\Big)$.

\subsection*{Sampling $\beta$}
Proceeding as before, when all $\pi_i$'s are zero, the corresponding $\beta_i$'s are all sampled from the Dirac Delta function $\delta_o$ resulting in all zeros. For non-zero vector $\mathbf{\pi}$ = $\{\pi_1, \ldots \pi_p\}$, the conditional posterior of $\mathbf{\beta}$ is obtained as follows:

\begin{align*} \label{eq:beta_post}
\begin{split}
p(\mathbf{\beta}|\mathbf{y}, \mathbf{\pi}, \sigma^2, \tau^2)  = 
 = \frac{1}{Z}exp\Big( -\frac{1}{2} \Big( \mathbf{\beta} - \Big( \mathbf{X}^T\mathbf{X} \frac{1}{\sigma^2} + \mathbf{I}\frac{1}{\sigma^2\tau^2}\Big)^{-1}\mathbf{X}^T \mathbf{y}\frac{1}{\sigma^2} \Big)^T \\ \Big( \mathbf{X}^T\mathbf{X}\frac{1}{\sigma^2} + \mathbf{I}\frac{1}{\sigma^2\tau^2}\Big) \Big( \mathbf{\beta} - \Big( \mathbf{X}^T\mathbf{X} \frac{1}{\sigma^2} + \mathbf{I}\frac{1}{\sigma^2\tau^2}\Big)^{-1}\mathbf{X}^T \mathbf{y}\frac{1}{\sigma^2} \Big)\Big)
\end{split}
\end{align*}

Since this is the kernel of a Gaussian distribution, we can now sample all $\beta_i$'s as follows

\begin{equation}
\beta_i | \mathbf{y}, \pi_i, \sigma^2, \tau^2 \sim  
\begin{cases}
\delta_0 \ \ \ \ \ \ \ \ \ \  \ \ \ \ \ \ \ \ \ \ \ \ \ \ \ \ \ \ \ \ \ \ \ \ \ \ \ \   ,\pi_i = 0 \\
N \Big( \Big( \mathbf{X}^T\mathbf{X} \frac{1}{\sigma^2} + \mathbf{I}\frac{1}{\sigma^2\tau^2}\Big)^{-1}\mathbf{X}^T \mathbf{y}\frac{1}{\sigma^2}, \\ \Big( \mathbf{X}^T\mathbf{X}\frac{1}{\sigma^2} + \mathbf{I}\frac{1}{\sigma^2\tau^2}\Big)^{-1}\Big) \  \  ,\pi_i = 1 \\
\end{cases}
\end{equation}

\subsection*{Sampling $\pi$}
The individual $\pi_j$'s are conditionally independent given $\theta$. We compare two cases: one when the $j^{th}$ element of $\mathbf{\beta}$ is zero or $\pi_j$ is zero and the other when $\pi_j$=1. We denote by $\mathbf{\pi}_{-j}$ the state of the variables barring $j$. Let $\pi_j=1|\mathbf{y}, \mathbf{\beta}_{-j}, \mathbf{\pi}_{-j}, \sigma^2, \tau^2, \theta$ $\sim \mbox{Bern}(\zeta_j)$. Let $a = p(\pi_j=1|\mathbf{y}, \mathbf{\beta}_{-j}, \mathbf{\pi}_{-j}, \sigma^2, \tau^2, \theta)$ and $b=\pi_j=1|\mathbf{y}, \mathbf{\beta}_{-j}, \mathbf{\pi}_{-j}, \sigma^2, \tau^2, \theta$. Then $\zeta_j = \frac{a}{a+b}$. We then draw $\pi_j$ from a Bernoulli with a chance parameter $\zeta_j$ and we repeat this for all predictors $\beta_j$. For the case when $\pi_j=0$,

\begin{align} \label{eq:pi_post_2}
\begin{split}
p(\pi_j=0|\mathbf{y}, \mathbf{\beta}_{-j}, \mathbf{\pi}_{-j}, \sigma^2, \tau^2, \theta)  \\
= \frac{1}{Z}exp\Big( - \frac{1}{2\sigma^2} (\mathbf{y} - \mathbf{X}_{-j} \mathbf{\beta}_{-j} )^T (\mathbf{y} - \mathbf{X}_{-j} \mathbf{\beta}_{-j} ) \Big) (1-\theta)
\end{split}
\end{align}

\noindent where we have absorbed all the irrelevant terms into $Z$, the normalizing constant. The expression for $\pi_j=1$ can be written similarly except that it would require integration over $\beta_j$. Defining $\mathbf{z}$ = $\mathbf{y} - \mathbf{X}_{-j} \beta_{-j}$, we have

\begin{align} \label{eq:pi_post_3}
\begin{split}
p(\pi_j=1|\mathbf{y}, \mathbf{\beta}_{-j}, \mathbf{\pi}_{-j}, \sigma^2, \tau^2, \theta)  \\
= \frac{1}{Z} \theta (2\pi \sigma^2 \tau^2)^{-\frac{1}{2}} exp\Big( -\frac{1}{2\sigma^2} (\mathbf{y} - \mathbf{X}_{-j} \mathbf{\beta}_{-j} )^T (\mathbf{y} - \mathbf{X}_{-j} \mathbf{\beta}_{-j} ) \Big) \\  exp\Big(  \frac{(\sum_{i=1}^n x_i z_i)^2}{2\sigma^2(\sum_{i=1}^n x_i^2 + \frac{1}{\tau^2})} \Big) \\
\end{split}
\end{align}

The conditional posterior of $\pi=0$ is therefore a Bernoulli distribution with chance parameter

\begin{align} \label{eq:pi_post_5}
\begin{split}
1-\zeta_j 
= \frac{1-\theta}{(\sigma^2\tau^2)^{-\frac{1}{2}} exp(K)\Big( \frac{\sigma^2}{(\sum_{i=1}^n x_i^2 + \frac{1}{\tau^2})} \Big)^{\frac{1}{2}}\theta + (1-\theta)}
\end{split}
\end{align}

\noindent where $K=\frac{(\sum_{i=1}^n x_i z_i)^2}{2\sigma^2(\sum_{i=1}^n x_i^2 + \frac{1}{tau^2})}$ and where $z_j$ changes depending on which $\beta_j$ we sample. 

\begin{table}[!t]
\small
\begin{tabular}{|l|l|l|l|l|l|l|}
\hline
\textbf{Sector Name}                           & \multicolumn{2}{c|}{\begin{tabular}[c]{@{}l@{}}\textbf{Treatment} \\ \textbf{ (RMSE)} \end{tabular} }                     & \multicolumn{2}{c|}{\begin{tabular}[c]{@{}l@{}} \textbf{Control} \\ \textbf{ (RMSE)}  \end{tabular}}                       & \multicolumn{2}{l|}{\textbf{p - score}}                    \\ \hline
\textbf{}                                      & \textbf{SSR}                  & \textbf{LR}                 & \textbf{SSR}                  & \textbf{LR}                 & \textbf{F1}                 & \textbf{Acc. \%}             \\ \hline
{\color[HTML]{000000} \textbf{Manufacturing}}  & {\color[HTML]{000000} 12.34} & {\color[HTML]{000000} 12.39} & {\color[HTML]{000000} 5.44}  & {\color[HTML]{000000} 5.38}  & {\color[HTML]{000000} 0.68} & {\color[HTML]{000000} 64.84} \\ \hline
{\color[HTML]{000000} \textbf{Transportation}} & {\color[HTML]{000000} 12.68} & {\color[HTML]{000000} 12.83} & {\color[HTML]{000000} 13.98} & {\color[HTML]{000000} 14.17} & {\color[HTML]{000000} 0.69} & {\color[HTML]{000000} 61.76} \\ \hline
{\color[HTML]{000000} \textbf{Clothing}}       &  {\color[HTML]{000000} 10.01} & {\color[HTML]{000000} 10.01} & {\color[HTML]{000000} 4.89}  & {\color[HTML]{000000} 5.02}  & {\color[HTML]{000000} 0.64} & {\color[HTML]{000000} 62.52} \\ \hline
{\color[HTML]{000000} \textbf{Personal Use}}   & {\color[HTML]{000000} 9.84}  & {\color[HTML]{000000} 10}    & {\color[HTML]{000000} 10.71} & {\color[HTML]{000000} 10.84} & {\color[HTML]{000000} 0.65} & {\color[HTML]{000000} 64.32} \\ \hline
{\color[HTML]{000000} \textbf{Housing}}        & {\color[HTML]{000000} 12.1}  & {\color[HTML]{000000} 12.23} & {\color[HTML]{000000} 11.79} & {\color[HTML]{000000} 11.93} & {\color[HTML]{000000} 0.68} & {\color[HTML]{000000} 60}    \\ \hline
{\color[HTML]{000000} \textbf{Food}}           & {\color[HTML]{000000} 11.42} & {\color[HTML]{000000} 11.61} & {\color[HTML]{000000} 10.17} & {\color[HTML]{000000} 10.36} & {\color[HTML]{000000} 0.69} & {\color[HTML]{000000} 66.96} \\ \hline
{\color[HTML]{000000} \textbf{Arts}}           & {\color[HTML]{000000} 11.11} & {\color[HTML]{000000} 11.21} & {\color[HTML]{000000} 12.03} & {\color[HTML]{000000} 12.11} & {\color[HTML]{000000} 0.7}  & {\color[HTML]{000000} 59.69} \\ \hline
{\color[HTML]{000000} \textbf{Retail}}         & {\color[HTML]{000000} 11.35} & {\color[HTML]{000000} 11.69} & {\color[HTML]{000000} 11.32} & {\color[HTML]{000000} 11.45} & {\color[HTML]{000000} 0.74} & {\color[HTML]{000000} 71.82} \\ \hline
{\color[HTML]{000000} \textbf{Construction}}   & {\color[HTML]{000000} 10.15} & {\color[HTML]{000000} 10.21} & {\color[HTML]{000000} 10.15} & {\color[HTML]{000000} 10.21} & {\color[HTML]{000000} 0.7}  & {\color[HTML]{000000} 69.33} \\ \hline
{\color[HTML]{000000} \textbf{Agriculture}}    & {\color[HTML]{000000} 10.66} & {\color[HTML]{000000} 10.75} & {\color[HTML]{000000} 11.39} & {\color[HTML]{000000} 11.55} & {\color[HTML]{000000} 0.71} & {\color[HTML]{000000} 62.89} \\ \hline
{\color[HTML]{000000} \textbf{Services}}       & {\color[HTML]{000000} 12.72} & {\color[HTML]{000000} 12.94} & {\color[HTML]{000000} 12.83} & {\color[HTML]{000000} 13}    & {\color[HTML]{000000} 0.74} & {\color[HTML]{000000} 68.35} \\ \hline
{\color[HTML]{000000} \textbf{Education}}      &  {\color[HTML]{000000} 12.57} & {\color[HTML]{000000} 12.67} & {\color[HTML]{000000} 6.18}  & {\color[HTML]{000000} 6.36}  & {\color[HTML]{000000} 0.66} & {\color[HTML]{000000} 61.05} \\ \hline
\end{tabular}
\caption{Result comparison on the test set. For p-score estimation, we use F1 score and accuracy (the higher, the better); for outcome estimations, we use RMSE (the lower the better). }
\vspace{-8mm}
\label{tab:regress_2}
\end{table}

\section{Experiments and Results} \label{sec:results_1}
In this section, we first start by evaluating the effectiveness of the learning methods in modeling individual estimators that form the components of $\hat{\tau}_{DRE}$. The outcome models through spike-and-slab Bayesian variable selection models have been described in the previous sections. For estimating the propensity score $e(x)$ = $\mathbb{P}(W=1|X=x)$ with estimator $\hat{e}(x)$ in step 2 outlined in Section~\ref{sec:steps_causal}, we use a logistic regression model with the same attributes as the outcome model. We further experimented with Random Forests, but did not observe any substantial difference in the results. For the Gibbs sampling procedure, we set the following hyper-parameter values: $a=b=1$, $a_1= a_2=0.01$, $\theta$=0.5 and $s=1/2$ for all the models. We use a burn-in of 1000 samples for the procedure and use 4000 samples for the sampling procedure. We use these posterior estimates as the coefficient estimates in the spike-and-slab regression model for predictive purposes.

As mentioned before, for each sector, we consider treated and control groups considering that sector and evaluate the outcome models for treatment and control and the propensity score (p-score) models. We use Root Mean Squared Error (RMSE) for the outcome regression models and F1 sore and Accuracy for the p-score model using logistic regression. For each model we split the data into 70\%-30\% train-test and evaluate the models using these metrics on the held-out test set. The results shown for all sectors in Table~\ref{tab:regress_2} compares Linear Regression (LR) without any regularization with the Spike and Slab (SSR) model. We find that while for most sectors the models fare comparably for both treated and control groups, for 3 sectors namely Manufacturing, Clothing and Education where the regression models for Treatment are an order of magnitude worse than control groups evidenced by their RMSE scores. This can be attributed to the relatively low number of projects in these areas shown in Figure~\ref{fig:loans_stats}. We also find that the SSR model outperforms the LR model in most cases in terms of lower RMSE scores for the SSR model. For the p-score model, we find that the logistic regression model performs similar for most sectors showing lesser disparity among the several models used for the purpose. 

\begin{table}[!t]
\small
\begin{tabular}{|l|l|l|l|l|l|l|}
\hline
\multicolumn{1}{|c|}{}                                       & \multicolumn{2}{c|}{\textbf{Naive}}                        & \multicolumn{2}{c|}{\textbf{Baseline}}                     & \multicolumn{2}{c|}{\textbf{DRE (SSR)}}                    \\ \cline{2-7} 
\multicolumn{1}{|c|}{{\textbf{Sector Name}}} & \textbf{ATE}                 & \textbf{std}                & \textbf{ATE}                 & \textbf{std}                & \textbf{ATE}                 & \textbf{std}                \\ \hline
{\color[HTML]{000000} \textbf{Construction}}                 & {\color[HTML]{000000} 1.04}  & {\color[HTML]{000000} 0.28} & {\color[HTML]{000000} -0.09} & {\color[HTML]{000000} 0.27} & {\color[HTML]{000000} 0.51}  & {\color[HTML]{000000} 0.29} \\ \hline
{\color[HTML]{000000} \textbf{Clothing}}                     & {\color[HTML]{000000} 1.85}  & {\color[HTML]{000000} 0.16} & {\color[HTML]{000000} 3.12}  & {\color[HTML]{000000} 0.15} & {\color[HTML]{000000} 2.63}  & {\color[HTML]{000000} 0.29} \\ \hline
{\color[HTML]{000000} \textbf{Retail}}                       & {\color[HTML]{000000} 0.59}  & {\color[HTML]{000000} 0.15} & {\color[HTML]{000000} 3.25}  & {\color[HTML]{000000} 0.15} & {\color[HTML]{FE0000} 4.81*}  & {\color[HTML]{000000} 0.18} \\ \hline
{\color[HTML]{000000} \textbf{Education}}                    & {\color[HTML]{000000} -4.86} & {\color[HTML]{000000} 0.2}  & {\color[HTML]{000000} -5.48} & {\color[HTML]{000000} 0.19} & {\color[HTML]{000000} -5.37} & {\color[HTML]{000000} 0.19} \\ \hline
{\color[HTML]{000000} \textbf{Services}}                     & {\color[HTML]{000000} -0.52} & {\color[HTML]{000000} 0.15} & {\color[HTML]{000000} -0.92} & {\color[HTML]{000000} 0.14} & {\color[HTML]{000000} -0.85} & {\color[HTML]{000000} 0.15} \\ \hline
{\color[HTML]{000000} \textbf{Manufacturing}}                & {\color[HTML]{000000} -5.16} & {\color[HTML]{000000} 0.25} & {\color[HTML]{000000} -5.53} & {\color[HTML]{000000} 0.23} & {\color[HTML]{000000} -5.38} & {\color[HTML]{000000} 0.24} \\ \hline
{\color[HTML]{000000} \textbf{Transportation}}               & {\color[HTML]{000000} 1.34}  & {\color[HTML]{000000} 0.22} & {\color[HTML]{000000} 1.6}   & {\color[HTML]{000000} 0.2}  & {\color[HTML]{000000} 1.05}  & {\color[HTML]{000000} 0.22} \\ \hline
{\color[HTML]{000000} \textbf{Agriculture}}                  & {\color[HTML]{000000} -0.61} & {\color[HTML]{000000} 0.16} & {\color[HTML]{000000} -0.28} & {\color[HTML]{000000} 0.15} & {\color[HTML]{000000} -0.6}  & {\color[HTML]{000000} 0.15} \\ \hline
{\color[HTML]{000000} \textbf{Housing}}                      & {\color[HTML]{000000} 6.34}  & {\color[HTML]{000000} 0.19} & {\color[HTML]{000000} 6.77}  & {\color[HTML]{000000} 0.18} & {\color[HTML]{FE0000} 7.9*}   & {\color[HTML]{000000} 0.22} \\ \hline
{\color[HTML]{000000} \textbf{Arts}}                         & {\color[HTML]{000000} -5.46} & {\color[HTML]{000000} 0.23} & {\color[HTML]{000000} -5.4}  & {\color[HTML]{000000} 0.22} & {\color[HTML]{FE0000} -5.61} & {\color[HTML]{000000} 0.26} \\ \hline
{\color[HTML]{000000} \textbf{Personal Use}}                 & {\color[HTML]{000000} 1.61}  & {\color[HTML]{000000} 0.27} & {\color[HTML]{000000} 1.35}  & {\color[HTML]{000000} 0.25} & {\color[HTML]{FE0000} -2.26*} & {\color[HTML]{000000} 0.71} \\ \hline
{\color[HTML]{000000} \textbf{Food}}                         & {\color[HTML]{000000} -1.43} & {\color[HTML]{000000} 0.15} & {\color[HTML]{000000} 0.34}  & {\color[HTML]{000000} 0.14} & {\color[HTML]{000000} -0.4}  & {\color[HTML]{000000} 0.17} \\ \hline
\end{tabular}
\caption{Summary of ATE Estimation for different sectors comparing models. Numbers marked in asterisk indicate substantial differences in the estimates from the regression coefficients estimated in Table~\ref{tab:regress_1}}.
\vspace{-10mm}
\label{tab:regress_3}
\end{table}

Next, we compare the ATE for different sectors against a model where the ATE is estimated with just the target variable - the funding time. We compare 3 models for measuring the Average Treatment Effect (ATE):
\begin{enumerate}
    \item \textbf{Naive} -  ATE is calculated using the differences in means of $Y$ for treatment and control groups, and the standard deviation is calculated using the group standard deviations.
    
    \item \textbf{Baseline} - Here we use the Linear Regression (LR) model as discussed above to estimate 2 relations: (1) $Y(1)$ = $\mathbf{X}\beta_1$ with estimate $\hat{\beta}_1$ using the treated data and $Y(0)$ = $\mathbf{X}\beta_0$ with estimate $\hat{\beta}_0$ using the control data. The estimator $\hat{\tau}$ = $\frac{1}{n}\sum_{i=1}^n (\hat{Y}_{1, i} - \hat{Y}_{0, i})$. The standard error is then calculated as $\sqrt{\frac{var(Y_i - \hat{Y}_{1,i} | i: W_i=1)}{n_t -1 } + \frac{var(Y_i - \hat{Y}_{0,i} | i: W_i=0)}{n_t -1 }}$. 
    
    \item \textbf{DRE (SSR)} - Here we use the SSR models for the estimators $\hat{Y}_1$ and $\hat{Y}_0$ from the treated and control data and $\hat{\tau}_{DRE}$ and teh standard errro is calculated as described in Section~\ref{sec:steps_causal}. 
\end{enumerate}

The results for the model is shown in Table~\ref{tab:regress_3}. From the table, we find that the four sectors where the ATE from the DRE estimator is substantially different from the naive estimator are Retail, Housing, Arts and Personal Use (we keep the 3 sectors, Manufacturing, clothing and education out of our discussion since the SSR models for the treated data in these 3 sectors were substantially worse than control data). In fact, we find that the funding time for Arts loans have almost 6 days (ATE=-5.61) faster funding when compared to all other sectors using our DRE (SSR) model, whereas the naive estimator suggests a slower funding. This suggests that when we combine these economic factors along with the loan attributes for these specific sectors, the effect of this loan sector actually helps in faster funding which in other situations would have been difficult to be funded. Similarly, for the Retail loans, we find that funding is generally disfavored compared to other factors by being funded slower by 5 (ATE=4.81) days. The standard errors for all the 3 models are comparable and so as such the ATE estimates can be compared reliably across the models. These observations suggest that when such economic disparities or similarities exist which can affect lender trust and perceptions of funding a project in a particular sector, biases are bound to arise. Therefore, predictive models which try to model the risk of loan defaults must also incorporate fairness constraints to not allow favoritism towards certain sectors. To this end, we conclude this study by modifying our SSR model to  incorporate fairness constraints.

\section{Controlling the disparities from sectors}
To control the disparities arising from the different attributes for different sectors in our regression setting, we adopt the procedure described in \cite{calders2013controlling} and incorporate the constraint in the sampling procedure for the parameter estimates. For each sector $s$, we divide the dataset as done before into two groups: $D_s^{\uparrow}$ and $D_s^{\downarrow}$ based on $s$. The specific goal here is to build one regression model for each sector and learn the parameters of that model while minimizing bias associated with predicting the target variables when conditioned over the loan sector attribute. To this end, we use the constraint that ensures that the mean predictions for the two groups $D_s^{\uparrow}$ and $D_s^{\downarrow}$ are equal irrespective of what the target or outcome exhibits. 

\subsection{Adding Regularization}
We use the same model based on Bayesian variable selection introduced in Section~\ref{sec:learning_model} with the addition of new regularization terms. We add the sector attribute to the features $\mathbf{X}$, however we now build one single model for each sector with the entire batch of data.  We use the balanced means constraint based on the following criteria: $\frac{\sum_{(\mathbf{x}_i, t_i) \in D_s^\uparrow} \mathbf{\beta}.\mathbf{x}_i }{|D_s^\uparrow|} = \frac{\sum_{(\mathbf{x}_i, t_i) \in D_s^\downarrow} \mathbf{\beta}.\mathbf{x}_i }{|D_s^\downarrow|}$, where $D_S^{\uparrow}$ and $D_S^{\downarrow}$ denote control and treatment data.
It denotes the constraint that the predictions from our model should be the same for both the treated and the control groups for the loan sector in consideration irrespective of what the target variable differences in the model exhibit.  Using the same notations used in Equations~\ref{eq:spike_slab}, we make the following adjustment to sample the target variable. Denoting $\frac{\sum_{(\mathbf{x}_i, t_i) \in D_s^\uparrow} \mathbf{\beta}.\mathbf{x}_i }{|D_s^\uparrow|} = \frac{\sum_{(\mathbf{x}_i, t_i) \in D_s^\downarrow} \mathbf{\beta}.\mathbf{x}_i }{|D_s^\downarrow|}$ as $\mathbf{d}$, we add the regularization term as: $y_i \sim N(\beta x_i, \sigma^2) + \lambda\mathbf{\beta}\mathbf{d}$, where $\lambda$ is the hyper-parameter controlling the effect of the regularization term.  With this modification, the sampling equations are modified in the following way: \\
\noindent \textbf{Sampling $\sigma^2$}: The posterior is now sampled from $
\sigma^2|\mathbf{y}, \mathbf{\beta} \sim \mbox{Gamma} \Big( \alpha_1 + \frac{n}{2} , \alpha_2 + 
\frac{(\mathbf{y} - \mathbf{X}\beta)^T(\mathbf{y} - \mathbf{X}\beta) + \lambda(\mathbf{\beta} \mathbf{d})}{2}\Big) $

\noindent \textbf{Sampling $\beta$}: The posterior can now be sampled from 

\begin{equation}
\beta_i | \mathbf{y}, \pi_i, \sigma^2, \tau^2 \sim  
\begin{cases}
\delta_0 \ \ \ \ \ \ \ \ \ \  \ \ \ \ \ \ \ \ \ \ \ \ \ \ \ \ \ \ \ \ \ \ \ \ \ \ \ \   ,\pi_i = 0 \\
N \Big( \Big( \mathbf{X}^T\mathbf{X} \frac{1}{\sigma^2} + \mathbf{I}\frac{1}{\sigma^2\tau^2}\Big)^{-1} \\ \frac{1}{\sigma^2}(\mathbf{X}^T\mathbf{y} - \mathbf{d}) \\ \Big( \mathbf{X}^T\mathbf{X}\frac{1}{\sigma^2} + \mathbf{I}\frac{1}{\sigma^2\tau^2}\Big)^{-1}\Big) \  \  ,\pi_i = 1 \\
\end{cases}
\end{equation}

\noindent \textbf{Sampling $\pi$}: The posterior for $\pi_j$ can similarly be sampled. For the case when $\pi_j=0$,

\begin{align} \label{eq:pi_post_fair_1}
\begin{split}
p(\pi_j=0|\mathbf{y}, \mathbf{\beta}_{-j}, \mathbf{\pi}_{-j}, \sigma^2, \tau^2, \theta)  \\
= \frac{1}{Z}exp\Big( - \frac{1}{2\sigma^2} (\mathbf{y} - \mathbf{X}_{-j} \mathbf{\beta}_{-j} )^T (\mathbf{y} - \mathbf{X}_{-j} \mathbf{\beta}_{-j} ) \\ + \lambda(\beta \mathbf{d}) \Big) (1-\theta)
\end{split}
\end{align}

As done before, we define $\mathbf{z}$ = $\mathbf{y} - \mathbf{X}_{-j} \beta_{-j}$ as the residuals of the regression $\mathbf{y}$ on $\mathbf{X}_{-j}$. Modifying Equation ~\ref{eq:pi_post_3}, we arrive at

\begin{align} \label{eq:pi_post_fair_3}
\begin{split}
p(\pi_j=1|\mathbf{y}, \mathbf{\beta}_{-j}, \mathbf{\pi}_{-j}, \sigma^2, \tau^2, \theta)  \\
= \frac{1}{Z} \theta (2\pi \sigma^2 \tau^2)^{-\frac{1}{2}} exp\Big( \\ -\frac{1}{2\sigma^2} (\mathbf{y} - \mathbf{X}_{-j} \mathbf{\beta}_{-j} )^T (\mathbf{y} - \mathbf{X}_{-j} \mathbf{\beta}_{-j} ) + \beta_{-j} \mathbf{d}_{-j} \Big) \\  exp\Big(  \frac{(\sum_{i=1}^n x_i z_i + d_i)^2}{2\sigma^2(\sum_{i=1}^n x_i^2 + \frac{1}{\tau^2})} \Big) \\
\end{split}
\end{align}

The conditional posterior of $\pi=0$ is therefore a Bernoulli distribution with chance parameter

\begin{align} \label{eq:pi_post_5}
\begin{split}
1-\zeta_j 
= \frac{1-\theta}{(\sigma^2\tau^2)^{-\frac{1}{2}} exp(K)\Big( \frac{\sigma^2}{(\sum_{i=1}^n x_i^2 + \frac{1}{\tau^2})} \Big)^{\frac{1}{2}}\theta + (1-\theta)}
\end{split}
\end{align}

\noindent where $K=\frac{(\sum_{i=1}^n x_i z_i + d+i)^2}{2\sigma^2(\sum_{i=1}^n x_i^2 + \frac{1}{tau^2})}$ and where $z_j$ changes depending on which $\beta_j$ we sample.

\subsection{Results}
Finally, we compare the results of the models with the regularization constraint for the sectors with models discussed prior to this. Additionally, we also compare the results from the model in the absence of external factors and only considering loan attributes available from Kiva data. We adopt a similar validation approach as previous where we peform a 70\%-30\% train-test split and test on the held-out 30\% data. For training the SSR models, we use the same settings as explained in Section~\ref{sec:results_1} for the Gibbs sampling procedure. For evaluating the regression models, we use the metric RMSE on the test data as done in the previous section. The regularization hyper-parameter $\lambda$, we set it to 0.6 after cross-validating it with several values. The results have been shown in Table~\ref{tab:regress_4} - the column LR-LA shows the results for the model with only loan attributes from Kiva. The last column shows results incorporating the regularization term. Additionally, we only test the models with the 4 sectors that showed the highest ATE explained in Section~\ref{sec:results_1}. 

We observe that in all these sectors,  addition of external factors like the economic attributes and borrower-lender country pair attributes improve the model over the model LR-LA. The model with SSR performs the best in the absence of any regularization for all the sectors having the least RMSE, indicating that variable selection helps improve the predictions. However, when we compare these results with the model SSR (with regularization), we find that the performance drops at the cost of the equality constraints, however what we observe is that the results are still comparable to the simple LR model. We find that for Housing loans, the model with regularization performs comparably worse and this can be attributed to the pre-existing disparities shown by high ATE for these loans as shown in Table~\ref{tab:regress_3}. Therefore,  the equal means constraint does result in performance degradation. However, these results suggest that we can still build models by reducing disparities in the resulting predictions while limiting the drop in performance.

\begin{table}[!t]
\small
\begin{tabular}{|l|l|l|l|l|}
\hline
\textbf{Sector}       & \textbf{LR} & \textbf{LR - LA} & \textbf{SSR} & \textbf{SSR (regularization)} \\ \hline
\textbf{Housing}      & 10.76       & 11.34            & 10.61        & 13.82                              \\ \hline
\textbf{Personal Use} & 9.6         & 10.02             & 9.46         & 10.24                            \\ \hline
\textbf{Retail}       & 12.06       & 13.18           & 11.91        & 12.73                              \\ \hline
\textbf{Arts}         & 9.31        & 10.25             & 9.19         & 9.52                               \\ \hline
\end{tabular}
\caption{RMSE results of regression models. Models with LA denote only loan attributes from Kiva are used in the model. The lower values indicate better results.} \label{tab:regress_4}
\vspace{-8mm}
\end{table}

\section{Related Work}

Understanding the effect of loan attributes towards  funding speeds have been studied extensively in \cite{ly2010individual} albeit only with factors from the loans data. The effects of cultural differences have also been studied in \cite{burtch2014cultural} where the authors present evidence that lenders prefer culturally similar borrowers in Kiva. However, the extent to which that affects the actual interests towards particular sectors was not presented. Our work here opens an entire body of research into fairness aware recommendation systems \cite{li2020fairness, berk2017convex} that might be necessary when promoting projects so as to lessen the inherent biases arising from existing lenders.  Especially when designing portfolio recommendations as a tool for decision support  for lenders as done in \cite{zhao2016portfolio}, it is important to adjust the multi-objective optimization problems incorporating constraints as described in this paper. Such conclusions can also be extended to platforms which are designed for lenders to profit from investments such as Lendingclub \cite{nowak2018small}. 

\section{Conclusions and Future Work}
In this paper, we first demonstrated how simple economic factors can play a role in deciding the speed of funding for particular loans and how they can be intertwined with the loan sector. We then measured the existing disparities arising from such factors using causal reasoning estimators and proposed a method to control the differences in outcome. One area where our work can be extended is to develop a single model taking all models into account - this is where the Bayesian variable selection method can be extended to incorporate priors that take into account fairness constraints for all sectors and using empirical bayes to drive the priors.
\bibliography{example_paper}
\bibliographystyle{ACM-Reference-Format}

\appendix
\subsection*{Sampling $\theta$}
 To sample the conditional posterior $p(\theta|\mathbf{y}, \mathbf{\beta}, \mathbf{\pi}, \tau^2, \sigma^2)$= $p(\theta|\mathbf{\pi})$, we calculate the following as:

\begin{align} \label{eq:theta_post}
\begin{split}
p(\theta|\mathbf{\pi}) &= \frac{p(\mathbf{\pi}|\theta)p(\theta)}{\int p(\mathbf{\pi}|\theta)p(\theta)d\theta} \\
                       &= \frac{\prod_i \theta^{\pi_i}(1-\theta)^{\pi_i} \frac{1}{B(a, b)}\theta^{a-1} (1-\theta)^{b-1}}{\int \prod_i \theta^{\pi_i}(1-\theta)^{\pi_i} \frac{1}{B(a, b)}\theta^{a-1} (1-\theta)^{b-1}d\theta} \\
                       &= \frac{\theta^{a+\sum_{i=1}^p \pi_i-1}(1-\theta)^{b+\sum_{i=1}^n(1-\pi_i)-1}}{\int \theta^{a+\sum_{i=1}^p \pi_i-1}(1-\theta)^{b+\sum_{i=1}^n(1-\pi_i)-1} d\theta}
\end{split}
\end{align}

where $B$ is the beta function and where we note that the numerator in Equation~\ref{eq:theta_post} is the kernel of a Beta distribution and the denominator is a normalizing constant. The posterior is $\theta|\mathbf{\pi}$ $\sim Beta\Big( a + \sum_{i=1}^p \pi_i, b + \sum_{i=1}^n (1-\pi_i) \Big)$.

\subsection*{Sampling $\tau^2$}
The conditional posterior of $\tau^2$ can be derived from the probability $p(\tau^2|\mathbf{y}, \mathbf{\beta}, \mathbf{\pi}, \theta, \sigma^2)$= $p(\tau^2|\mathbf{\pi}, \mathbf{\beta})$. Proceeding as before

\begin{align} \label{eq:tau_post}
\begin{split}
p(\tau^2|\mathbf{\pi}, \mathbf{\beta}) &= \frac{p(\mathbf{\beta}|\tau^2, \mathbf{\pi})p(\pi)p(\tau^2)}{\int p(\mathbf{\beta}|\tau^2, \pi)p(\mathbf{\pi})p(\tau^2)d\tau^2} \\
                       &= \frac{p(\mathbf{\beta}|\tau^2, \pi)p(\tau^2)}{\int p(\mathbf{\beta}|\tau^2, \pi)p(\tau^2)d\tau^2} \\
\end{split}
\end{align}

Here since $\pi$ can assume values 0 or 1 we tackle each case independently and derive the following. We sample from the prior if all $\pi_i$'s are zero. Let $\mathbf{\pi}$=$\{\pi_1, \ldots, \pi_p\}$ be the vector of mixture weights and let $\mathbf{0}$ be a vector of zeros of length $p$. Following this, we have

\begin{align} \label{eq:tau_post_2}
\begin{split}
p(\tau^2|\mathbf{\pi}, \mathbf{\beta}) &= \frac{1}{Z}p(\mathbf{\beta}|\tau^2, \mathbf{\pi})p(\pi)p(\tau^2) \\
                       &= \frac{1}{Z} \prod_{i=1}^p\pi_i (2\pi\sigma^2\tau^2)^{-\frac{1}{2}} exp\Big( -\frac{1}{2\sigma^2\tau^2} \mathbf{\beta}^T\mathbf{\beta} \Big) \\ & \frac{\Big( \frac{s^2}{2}\Big)^{\frac{1}{2}}}{\Gamma\Big( \frac{1}{2}\Big)} (\tau^2)^{-\frac{1}{2}-1} exp\Big(- \frac{\frac{s^2}{2}}{\tau^2}\Big) \\
                       &=\frac{1}{Z} (\tau^2)^{-(\frac{1}{2} + \frac{\sum_{i=1}^p \pi_i}{2})-1}exp \Big( -\frac{(\frac{s^2}{2} + \frac{\mathbf{\beta}^T\mathbf{\beta}}{2\sigma^2})}{\tau^2}\Big)
\end{split}
\end{align}
which is a Gamma distribution and therefore we sample $\tau^2|\mathbf{\beta}, \mathbf{\pi}$ $\sim \mbox{Inverse-Gamma}(\frac{1}{2} + \frac{\sum_{i=1}^p\pi_i}{2}, \frac{s^2}{2} + \frac{\mathbf{\beta}^T\mathbf{\beta}}{2\sigma^2})$. On the other hand, when $\pi=0$, the $\beta_i$'s are 0 and we simply sample from the prior $\tau^2|\mathbf{\beta}, \mathbf{\pi}$ $\sim \mbox{Inverse-Gamma}(\frac{1}{2}, \frac{s^2}{2})$.

\subsection*{Sampling $\sigma^2$}
The conditional posterior of $\tau^2$ can be derived in a similar manner as above from the probability $p(\sigma^2|\mathbf{y}, \mathbf{\beta}, \mathbf{\pi}, \theta, \sigma^2)$= $p(\sigma^2|\mathbf{y}, \mathbf{\beta})$. Proceeding as before, we expand

\begin{align} \label{eq:sigma_post}
\begin{split}
p(\sigma^2|\mathbf{y}, \mathbf{\beta}) = \frac{1}{Z}p(\mathbf{y}|\sigma^2, \mathbf{\beta})p(\beta)p(\sigma^2) \\
                       = \frac{1}{Z} (2\pi\sigma^2)^{-\frac{n}{2}}exp\Big( -\frac{1}{2\sigma^2} \ (\mathbf{y} - \mathbf{X}\beta)^T(\mathbf{y} - \mathbf{X}\beta) \Big) \\  \frac{\alpha_2^{\alpha_1}}{\Gamma(\alpha_1)} (\sigma^2)^{-a_1-1}exp\Big(-\frac{\alpha_2}{\sigma^2}\Big)
\end{split}
\end{align}

which after reductions turn out to be an inverse Gamma distribution. So $\sigma^2|\mathbf{y}, \mathbf{\beta}$ $\sim \mbox{Gamma} \Big( \alpha_1 + \frac{n}{2} , \alpha_2 + \frac{(\mathbf{y} - \mathbf{X}\beta)^T(\mathbf{y} - \mathbf{X}\beta)}{2}\Big)$.

\subsection*{Sampling $\beta$}
Proceeding as before, when all $\pi_i$'s are zero, the corresponding $\beta_i$'s are all sampled from the Dirac Delta function $\delta_o$ resulting in all zeros. For non-zero vector $\mathbf{\pi}$ = $\{\pi_1, \ldots \pi_p\}$, the conditional posterior of $\mathbf{\beta}$ is obtained as follows:

\begin{align} \label{eq:beta_post}
\begin{split}
p(\mathbf{\beta}|\mathbf{y}, \mathbf{\pi}, \sigma^2, \tau^2) \\ = \frac{p(\mathbf{y}|\mathbf{\beta}, \mathbf{\pi}, \tau^2)p(\mathbf{\beta}|\mathbf{\pi}, \tau^2)p(\mathbf{\pi})p(\tau^2)}{\int p(\mathbf{y}|\mathbf{\beta}, \mathbf{\pi}, \tau^2)p(\mathbf{\beta}|\mathbf{\pi}, \tau^2)p(\mathbf{\pi})p(\tau^2)d\mathbf{\beta}} \\
=\frac{p(\mathbf{y}|\mathbf{\beta}, \mathbf{\pi}, \tau^2)p(\mathbf{\beta}|\mathbf{\pi}, \tau^2)}{\int p(\mathbf{y}|\mathbf{\beta}, \mathbf{\pi}, \tau^2)p(\mathbf{\beta}|\mathbf{\pi}, \tau^2)d\mathbf{\beta}}
\end{split}
\end{align}

Expanding as done before by introducing the normalizing constant, we get

\begin{align} \label{eq:beta_post_2}
\begin{split}
p(\mathbf{\beta}|\mathbf{y}, \mathbf{\pi}, \sigma^2, \tau^2)  \\ = \frac{1}{Z}(2\pi\sigma^2)^{-\frac{n}{2}}exp\Big(-\frac{1}{2\sigma^2} (\mathbf{y} - \mathbf{X\beta})^T(\mathbf{y} - \mathbf{X\beta})\Big)  (2\pi\sigma^2\tau^2)^{-\frac{1}{2}} \\ exp\Big(-\frac{1}{2\sigma^2\tau^2}\beta^2\Big) \\
= \frac{1}{Z}exp\Big(-\frac{1}{2\sigma^2}(\mathbf{y} - \mathbf{X\beta})^T(\mathbf{y} - \mathbf{X\beta}) - \frac{1}{2\sigma^2\tau^2}\beta^2 \Big) \\
= \frac{1}{Z}exp\Big(-\frac{1}{2\sigma^2}[\mathbf{y}^T \mathbf{y} - 2\mathbf{\beta}^T\mathbf{X}^T\mathbf{y} + \mathbf{\beta}^T\mathbf{X}^T\mathbf{X}\mathbf{\beta}] - \frac{1}{2\sigma^2\tau^2}\beta^2 \Big) \\
= \frac{1}{Z}exp\Big(-\frac{1}{2}\Big[ - 2\mathbf{\beta}^T\mathbf{X}^T\mathbf{y}\frac{1}{\sigma^2} + \mathbf{\beta}^T\mathbf{X}^T\mathbf{X}\frac{1}{\sigma^2}\mathbf{\beta} + \frac{1}{\sigma^2\tau^2}\beta^2 \Big] \Big) \\
= \frac{1}{Z}exp\Big(-\frac{1}{2}\Big[ \mathbf{\beta}^T\Big( \mathbf{X}^T\mathbf{X}\frac{1}{\sigma^2} + \mathbf{I}\frac{1}{\sigma^2\tau^2} \Big)\mathbf{\beta} - 2\mathbf{\beta}^T\mathbf{X}^T\mathbf{y} \frac{1}{\sigma^2} \Big] \Big)\\
\end{split}
\end{align}



Completing the squares trick leads us to 

\begin{align} \label{eq:beta_post_4}
\begin{split}
p(\mathbf{\beta}|\mathbf{y}, \mathbf{\pi}, \sigma^2, \tau^2)  \\ = \frac{1}{Z}exp\Big( -\frac{1}{2} \Big( \mathbf{\beta} - \Big( \mathbf{X}^T\mathbf{X} \frac{1}{\sigma^2} + \mathbf{I}\frac{1}{\sigma^2\tau^2}\Big)^{-1}\mathbf{X}^T \mathbf{y}\frac{1}{\sigma^2} \Big)^T \\ \Big( \mathbf{X}^T\mathbf{X}\frac{1}{\sigma^2} + \mathbf{I}\frac{1}{\sigma^2\tau^2}\Big) \Big( \mathbf{\beta} - \Big( \mathbf{X}^T\mathbf{X} \frac{1}{\sigma^2} + \mathbf{I}\frac{1}{\sigma^2\tau^2}\Big)^{-1}\mathbf{X}^T \mathbf{y}\frac{1}{\sigma^2} \Big)\Big)
\end{split}
\end{align}

Since this is the kernel of a Gaussian distribution, we can now sample all $\beta_i$'s as follows

\begin{equation}
\beta_i | \mathbf{y}, \pi_i, \sigma^2, \tau^2 \sim  
\begin{cases}
\delta_0 \ \ \ \ \ \ \ \ \ \  \ \ \ \ \ \ \ \ \ \ \ \ \ \ \ \ \ \ \ \ \ \ \ \ \ \ \ \   ,\pi_i = 0 \\
N \Big( \Big( \mathbf{X}^T\mathbf{X} \frac{1}{\sigma^2} + \mathbf{I}\frac{1}{\sigma^2\tau^2}\Big)^{-1}\mathbf{X}^T \mathbf{y}\frac{1}{\sigma^2}, \\ \Big( \mathbf{X}^T\mathbf{X}\frac{1}{\sigma^2} + \mathbf{I}\frac{1}{\sigma^2\tau^2}\Big)^{-1}\Big) \  \  ,\pi_i = 1 \\
\end{cases}
\end{equation}

\subsection*{Sampling $\pi$}
The individual $\pi_j$'s are conditionally independent given $\theta$. We compare two cases: one when the $j^{th}$ element of $\mathbf{\beta}$ is zero or $\pi_j$ is zero and the other when $\pi_j$=1. We denote by $\mathbf{\pi}_{-j}$ the state of the variables barring $j$. Let $\pi_j=1|\mathbf{y}, \mathbf{\beta}_{-j}, \mathbf{\pi}_{-j}, \sigma^2, \tau^2, \theta$ $\sim \mbox{ Bern}(\zeta_j)$. Let $a = p(\pi_j=1|\mathbf{y}, \mathbf{\beta}_{-j}, \mathbf{\pi}_{-j}, \sigma^2, \tau^2, \theta)$ and $b=\pi_j=1|\mathbf{y}, \mathbf{\beta}_{-j}, \mathbf{\pi}_{-j}, \sigma^2, \tau^2, \theta$. Then 

\begin{align} \label{eq:pi_post_1}
\zeta_j = \frac{a}{a+b}
\end{align}

We then draw $\pi_j$ from a Bernoulli with a chance parameter $\zeta_j$ and we repeat this for all predictors $\beta_j$. For the case when $\pi_i=j=0$,

\begin{align} \label{eq:pi_post_2}
\begin{split}
p(\pi_j=0|\mathbf{y}, \mathbf{\beta}_{-j}, \mathbf{\pi}_{-j}, \sigma^2, \tau^2, \theta)  \\ = \frac{1}{Z} p(\mathbf{y} | \pi_j=0, \mathbf{\pi}_{-j}, \mathbf{\beta}_{-j}, \sigma^2, \tau^2, \theta) p(\beta_{-j}|\mathbf{\pi}_{-j}, \sigma^2, \tau^2, \theta) \\ p(\pi|\theta)p(\theta)p(\tau^2)p(\sigma^2) \\
= \frac{1}{Z} p(\mathbf{y} | \pi_j=0, \mathbf{\pi}_{-j}, \mathbf{\beta}_{-j}, \sigma^2, \tau^2, \theta)  p(\pi|\theta) \\
= \frac{1}{Z}exp\Big( - \frac{1}{2\sigma^2} (\mathbf{y} - \mathbf{X}_{-j} \mathbf{\beta}_{-j} )^T (\mathbf{y} - \mathbf{X}_{-j} \mathbf{\beta}_{-j} ) \Big) (1-\theta)
\end{split}
\end{align}

where we have absorbed all the irrelevant terms into $Z$, the normalizing constant. The expression for $\pi_j=1$ can be written similarly except that it would require integration over $\beta_j$. We have

\begin{align} \label{eq:pi_post_3}
\begin{split}
p(\pi_j=1|\mathbf{y}, \mathbf{\beta}_{-j}, \mathbf{\pi}_{-j}, \sigma^2, \tau^2, \theta)  \\
= \frac{1}{Z} p(\mathbf{y} | \pi_j=1, \mathbf{\pi}_{-j}, \mathbf{\beta}_{-j}, \sigma^2, \tau^2, \theta)  p(\pi|\theta) \\
= \frac{1}{Z} \int p(\mathbf{y} | \pi_j=1, \mathbf{\pi}_{-j}, \mathbf{\beta}_{-j}, \sigma^2, \tau^2, \theta)  p(\pi|\theta) d\beta_j \\
= \frac{1}{Z} \theta (2\pi \sigma^2 \tau^2)^{-\frac{1}{2}} \int exp\Big( - \frac{1}{2\sigma^2} (\mathbf{y} - \mathbf{X} \mathbf{\beta} )^T (\mathbf{y} - \mathbf{X} \mathbf{\beta} ) \\ - \frac{1}{2 \sigma^2 \tau^2}\beta_j^2 \Big) d\beta_j \\
\end{split}
\end{align}

To simplify the calculation, we define $\mathbf{Z}$ = $\mathbf{y} - \mathbf{X}_{-j} \beta_{-j}$ as the residuals of the regression $\mathbf{y}$ on $\mathbf{X}_{-j}$. Therefore Equation ~\ref{eq:pi_post_3} can be simplified to 

\begin{align} \label{eq:pi_post_4}
\begin{split}
p(\pi_j=1|\mathbf{y}, \mathbf{\beta}_{-j}, \mathbf{\pi}_{-j}, \sigma^2, \tau^2, \theta)  \\
= \frac{1}{Z} \theta (2\pi \sigma^2 \tau^2)^{-\frac{1}{2}} \int exp\Big( - \frac{1}{2\sigma^2} \sum_{i=1}^n(z_i - \beta_j x_i)^2 \\ - \frac{1}{2 \sigma^2 \tau^2}\beta_j^2 \Big) d\beta_j \\
= \frac{1}{Z} \theta (2\pi \sigma^2 \tau^2)^{-\frac{1}{2}} exp\Big( -\frac{1}{2\sigma^2} \mathbf{z}^T\mathbf{z} \Big) \\ \int exp\Big( - \frac{1}{2\sigma^2} \Big[ -2\beta_j\sum_{i=1}^n z_ix_i + \beta_j^2\sum_{i=1}^n x_i^2 \Big] - \\\frac{1}{2 \sigma^2 \tau^2}\beta_j^2 \Big) d\beta_j \\
\end{split}
\end{align}

On completing the squares for obtaining a Gaussian kernel, we have

\begin{align} \label{eq:pi_post_5}
\begin{split}
p(\pi_j=1|\mathbf{y}, \mathbf{\beta}_{-j}, \mathbf{\pi}_{-j}, \sigma^2, \tau^2, \theta)  \\
= \frac{1}{Z} \theta (2\pi \sigma^2 \tau^2)^{-\frac{1}{2}} exp\Big( \\ -\frac{1}{2\sigma^2} (\mathbf{y} - \mathbf{X}_{-j} \mathbf{\beta}_{-j} )^T (\mathbf{y} - \mathbf{X}_{-j} \mathbf{\beta}_{-j} ) \Big) \\  exp\Big(  \frac{(\sum_{i=1}^n x_i z_i)^2}{2\sigma^2(\sum_{i=1}^n x_i^2 + \frac{1}{\tau^2})} \Big) \\
\end{split}
\end{align}

The conditional posterior of $\pi=0$ is therefore a Bernoulli distribution with chance parameter

\begin{align} \label{eq:pi_post_5}
\begin{split}
1-\zeta_j 
= \frac{1-\theta}{(\sigma^2\tau^2)^{-\frac{1}{2}} exp(K)\Big( \frac{\sigma^2}{(\sum_{i=1}^n x_i^2 + \frac{1}{\tau^2})} \Big)^{\frac{1}{2}}\theta + (1-\theta)}
\end{split}
\end{align}

where $K=\frac{(\sum_{i=1}^n x_i z_i)^2}{2\sigma^2(\sum_{i=1}^n x_i^2 + \frac{1}{tau^2})}$ and where $z_j$ changes depending on which $\beta_j$ we sample.

\end{document}